\documentclass[aps,pra,superscriptaddress,twocolumn,showpacs]{revtex4}

\renewcommand*{\[}{\begin{equation}}

\renewcommand*{\]}{\end{equation}}

\def\PRA{{Phys.~Rev.~A} }

\def\PRL{{Phys.~Rev.~Lett.} }

\newcommand{\myscaleboxa}[1]{\scalebox{0.8}[0.95]{#1}}
\newcommand{\myscaleboxb}[1]{\scalebox{1.35}[1.35]{#1}}
\newcommand{\myscaleboxc}[1]{\scalebox{1.1}[1.2]{#1}}
\newcommand{\myscaleboxd}[1]{\scalebox{1.2}[1.2]{#1}}

\usepackage{epsfig}
\usepackage{bm}

\usepackage{hyperref}

\begin{document}

\title{New determination of structure parameters in strong field tunneling ionization theory of molecules}

\author{Song-Feng Zhao}

\affiliation{J. R. Macdonald Laboratory, Physics Department, Kansas
State University, Manhattan, Kansas 66506-2604, USA}

\affiliation{College of Physics and Electronic Engineering,
Northwest Normal University, Lanzhou, Gansu 730070, People's
Republic of China}

\author{Cheng Jin}
\affiliation{J. R. Macdonald Laboratory, Physics Department, Kansas
State University, Manhattan, Kansas 66506-2604, USA}
\affiliation{College of Physics and Electronic Engineering,
Northwest Normal University, Lanzhou, Gansu 730070, People's
Republic of China}

\author{Anh-Thu Le}
\affiliation{J. R. Macdonald Laboratory, Physics Department, Kansas
State University, Manhattan, Kansas 66506-2604, USA}

\author{T. F. Jiang}
\affiliation{J. R. Macdonald Laboratory, Physics Department, Kansas
State University, Manhattan, Kansas 66506-2604, USA}
\affiliation{Institute of Physics, National
 Chiao-Tung University, Hsinchu 30010 Taiwan}

\author{C. D. Lin}
\affiliation{J. R. Macdonald Laboratory, Physics Department, Kansas
State University, Manhattan, Kansas 66506-2604, USA}

\date{\today}

\begin{abstract}
In the strong field molecular tunneling ionization theory of Tong
{\it et al.} [Phys. Rev. A {\bf 66}, 033402 (2002)], the ionization
rate depends on the asymptotic wavefunction of the molecular orbital
from which the electron is removed. The orbital wavefunctions
obtained from standard quantum chemistry packages in general are not
good enough in the asymptotic region. Here we construct a
one-electron model potential for several linear molecules using
density functional theory (DFT). We show that the asymptotic
wavefunction can be improved with an iteration method and after one
iteration accurate asymptotic wavefunctions and structure parameters
are determined. With the new parameters we examine the
alignment-dependent tunneling ionization probabilities for several
molecules and compare with other calculations and with recent
measurements, including ionization from inner molecular orbitals.
\end{abstract}


\maketitle

\section{INTRODUCTION}
Tunneling ionization of molecules in strong infrared fields is the first step in many interesting strong-field phenomena such as high-order
harmonic generation (HHG), emission of high-energy above-threshold ionization (HATI) electrons and non-sequential double ionization (NSDI).
Essential understanding to these processes is the angle-dependent ionization probability $P(\theta)$ for a molecule fixed in space, where $\theta$
is the angle between the molecular axis and the polarization direction of the laser's electric field. Since molecules are generally not fixed in
space, i.e., not at a fixed alignment and/or orientation, experimental determination  of $P(\theta)$ from partially aligned molecules requires
additional assumptions. Alnaser {\it et al.} \cite{Alnaser2004} first determined $P(\theta)$ from NSDI processes where the alignment of the
molecule is determined by Coulomb explosion of the molecular ions. $P(\theta)$ can also be determined by ionizing partially aligned molecules
\cite{NRC03,Pavicic2007}, or by measuring the angular distribution of electrons removed by a circularly polarized laser
\cite{Akagi2009,Staudte2009}. In both methods the alignment of the molecular axis is determined by Coulomb explosion when the molecular ion is
further ionized by an intense circularly polarized laser. In all of these measurements, the $P(\theta)$ is not determined directly for a fixed
angle and some approximations are used in order to determine the alignment-dependent ionization probability.

Theoretically, $P(\theta)$ can in principle be obtained directly from numerical solution of the time-dependent Schr\"{o}dinger equation (TDSE).
However, even for the simplest H$_{2}^{+}$, the $P(\theta)$ obtained from solving TDSE by different groups still exhibits relatively large
differences. While calculations  of $P(\theta)$ for interesting multielectron molecular systems have been carried out using the time-dependent
density-functional theory (TDDFT) (see, for example, \cite{Son2009}), the accuracy of these calculations is difficult to evaluate. Furthermore,
these calculations are rather time-consuming. Beside these {\it ab initio} approaches, alignment-dependent tunneling ionization rate for molecules
can be calculated using simple models such as the molecular strong field approximation (SFA) \cite{Faisal2000,Kjeldsen2004}, or the molecular
tunneling ionization theory \cite{Tong2002}. The latter is the simplest and is a generalization  of the tunneling model of Ammosov, Delone and
Krainov (ADK) \cite{ADK1986} for atoms. In the molecular tunneling ionization model (MO-ADK) of Tong {\it et al.} \cite{Tong2002}, the ionization
rate for a molecule aligned at an angle $\theta$ with respect to the laser polarization axis is given analytically. The ionization rate depends on
the instantaneous electric field of the laser, the ionization potential of the molecule and some structure parameters of the orbital wavefunction
in the asymptotic region. Subsequent further extension of
   the MO-ADK theory can be found in
\cite{Zhao06,Brabec05,Fabrikant09}.

In Tong {\it et al.} \cite{Tong2002}, the structure parameters are extracted from molecular wavefunctions calculated using the multiple scattering
method \cite{Dill1974}. However, these days molecular wavefunctions are more easily accessible from quantum chemistry packages such as GAMESS
\cite{Schmidt1993}, GAUSSIAN \cite{Frisch2003} and others. Thus it is desirable to obtain structure parameters from the asymptotic behavior of
orbitals calculated from such packages. This was carried out for CO$_{2}$ by Le {\it et al.} \cite{ATLe2007} and for other molecules by Kjeldsen
and Madsen \cite{Kjeldsen2005}. Unfortunately, molecular orbitals from these chemistry packages are calculated using gaussian basis functions and
they are not suitable for representing the exponential decay of the wavefunction at large distances. As more accurate experimental data are
becoming available, it is essential to redetermine  these structure parameters more accurately. Since the asymptotic wavefunction does not
contribute much to the total energy of a molecule, one cannot efficiently improve the asymptotic wavefunctions by enlarging the size of the
gaussian basis directly.

In this paper, we describe how to improve the asymptotic
wavefunction where the structure parameters are extracted. Our input
consists of wavefunctions of all the occupied orbitals obtained from
GAMESS or GAUSSIAN. We then construct a single-active-electron model
potential and solve the time-{\em in}dependent Schr\"{o}dinger
equation to obtain the molecular orbital wavefunctions by an
iterative procedure. The details of the method are given in Section
II. We then apply the method to redetermine all the structure
parameters previously published in \cite{Tong2002}, and adding
structure parameters for some inner orbitals. We also determine the
structure parameters for a number of systems that have been
investigated experimentally. Using these new structure parameters we
examined the alignment dependence of ionization probabilities for
several systems. In most cases, the new results do not differ much
from what were presented in Tong {\it et al.} \cite{Tong2002}.
However, there are differences in some molecules. The strong
deviation in CO$_{2}$ has been reported recently \cite{Zhao2009}.

\section{THEORETICAL METHODS}
The theory part is divided into three subsections. We first present the method of generating a single-active-electron model potential for linear
molecules. We then discuss how to calculate the wavefunctions by solving the time-{\em in}dependent Schr\"{o}dinger equation with B-spline basis
functions. We will also briefly describe how to extract the structure parameters   in the MO-ADK theory.

\subsection{Construction of single-active-electron model potentials for linear molecules}

Single-active-electron model potential approach has been widely used
for describing atoms in strong-field physics (see, for example,
\cite{Tong2005}). This approach has also been used for molecular
targets recently \cite{samha2009,Zhao2009}. The one-electron model
potential consists of two parts: electrostatic and
exchange-correlation terms. It is well-known that the traditional
local-density approximation (LDA) for the exchange-correlation
potential does not give the correct ($-1/r$) potential in the
asymptotic region where the structure parameters are to be
extracted. In this paper, we follow Abu-samha and Madsen
\cite{samha2009} and use the LB potential, proposed by Leeuwen and
Baerends \cite{Leeuwen1994}, which will give the correct asymptotic
$-1/r$ behavior for neutral atoms and molecules. We note that a
similar LB potential, called LB$\alpha$ \cite{Schipper2000}, has
also been used by Chu and collaborators in their TDDFT approach
\cite{Chu2005,Son2009}.

For linear molecules, the model potential can be expressed in
single-center expansion as
\begin{equation}
V(r,\theta)=\sum_{l=0}^{l_{max}} v_{l}(r)P_{l}(\cos\theta).
\end{equation}
Here, $v_{l}$(r) is the radial component of the model potential and $P_{l}(cos\theta)$   the Legendre polynomial. Typically we choose
$l_{max}=40$. The radial potential is given by
\begin{equation}
v_{l}(r)=v_{l}^{nuc}(r)+v_{l}^{el}(r)+v_{l}^{ex}(r),
\end{equation}
where the first two terms represent  the electrostatic potential and the last term is the exchange interaction.

The electron-nucleus interaction $v_{l}^{nuc}(r)$ can be written as
\begin{equation}
v_{l}^{nuc}(r)=\sum_{i=1}^{N_{a}}v_{l}^{i}(r),
\end{equation}
where $i$ runs over the $N_{a}$ atoms in the molecule. Without loss of generality,
we assume that linear molecules are aligned along the $z$-axis,
then $v_{l}^{i}(r)$ can be expressed as
\begin{equation}
v_{l}^{i}(r)=\left\{
\begin{array}{ll}
-(\frac{r_{<}^{i}}{r_{>}^{i}})^{l}\frac{Z_{c}^{i}}{r_{>}^{i}} &
\quad \mbox{for
$z_{i}>0$} \\
-(-1)^{l}(\frac{r_{<}^{i}}{r_{>}^{i}})^{l}\frac{Z_{c}^{i}}{r_{>}^{i}}
& \quad \mbox{for $z_{i}<0$}
\end{array} \right.
\end{equation}
with $r_{<}^{i}$=min(r,$|z_{i}|$), $r_{>}^{i}$=max(r,$|z_{i}|$).
Here $Z_{c}^{i}$ and $z_{i}$ are the nuclear charge and the $z$
coordinate of the $i$th atom, respectively.

The partial Hartree potential $v_{l}^{el}(r)$ is given by
\begin{equation}
v_{l}^{el}(r)=\frac{4\pi}{2l+1}\int_{0}^{\infty}a_{l}(r')r'^{2}\frac{r_{<}^{l}}{r_{>}^{l+1}}dr'
\end{equation}
with $r_{<}$=min(r,$r'$), $r_{>}$=max(r,$r'$). Here
$a_{l}(r')$ is
\begin{equation}
a_{l}(r')=\frac{2l+1}{2}\int_{-1}^{1}\rho(r',\theta')P_{l}(\cos\theta')d(\cos\theta'),
\end{equation}
where $\rho$ is the total electron density in the molecule and
\begin{equation}
\rho(r',\theta')=\sum_{i=1}^{N_{e}}\frac{1}{2\pi}\int_{0}^{2\pi}|\Psi_{i}(r',\theta',\varphi')|^{2}d\varphi'.
\end{equation}
Here $i$ runs over all the $N_{e}$ electrons in the molecule. The
wavefunction of each molecular orbital can be obtained from quantum
chemistry packages such as GAMESS \cite{Schmidt1993} or GAUSSIAN
\cite{Frisch2003}.

For the partial exchange potential, it is written as
\begin{equation}
v_{l}^{ex}(r)=\frac{2l+1}{2}\int_{-1}^{1}V_{ex,\sigma}(r,\theta)P_{l}(\cos\theta)d(\cos\theta),
\end{equation}
where
\begin{equation}
V_{ex,\sigma}(r,\theta)=\alpha V_{ex,\sigma}^{LDA}(r,\theta)+V_{ex,\sigma}^{GC}(r,\theta)
\end{equation}
Here $V_{ex,\sigma}^{LDA}(r,\theta)$ is the LDA potential for  an electron with spin $\sigma$
\begin{equation}
V_{ex,\sigma}^{LDA}(r,\theta)=-\big[\frac{6}{\pi}\rho_{\sigma}(r,\theta)\big]^{1/3},
\end{equation}
where
\begin{equation}
\rho_{\sigma}(r,\theta)=\sum_{i=1}^{N_{\sigma}}\frac{1}{2\pi}\int_{0}^{2\pi}| \Psi_{i\sigma}(r,\theta,\varphi)|^{2}d\varphi.
\end{equation}
Here $i$ runs over the $N_{\sigma}$ electrons that have the same spin
as the active electron. The gradient correction term is given by \cite{Leeuwen1994}
\begin{equation}
V_{ex,\sigma}^{GC}(r,\theta)=-\frac{\beta\chi_{\sigma}^{2}(r,\theta)\rho_{\sigma}^{1/3}(r,\theta)}
{1+3\beta\chi_{\sigma}(r,\theta)\sinh^{-1}(\chi_{\sigma}(r,\theta))},
\end{equation}
where
$\chi_{\sigma}(r,\theta)=|\nabla\rho_{\sigma}(r,\theta)|\rho_{\sigma}^{-4/3}(r,\theta)$.
The parameters $\alpha$ and $\beta$ are chosen to be $1.0$ and
$0.05$, respectively throughout this paper. We note that for more
accurate binding energies, the correlation potential should be
included into Eq.~(9). In the so-called LB$\alpha$ model, the two
parameters $\alpha$ and $\beta$ are usually chosen to be $1.19$ and
$0.01$, respectively (see, \cite{Schipper2000}).

\subsection{Calculation of molecular wavefunctions by solving the time-{\em in}dependent Schr\"{o}dinger equation}
With the model potential constructed in the previous subsection, the
wavefunction for the active electron in a linear molecule can be
obtained by solving the following time-{\em in}dependent
Schr\"{o}dinger equation
\begin{equation}
 H_{el}\psi_{n}^{(m)}(\bm{r})\equiv[-\frac{1}{2}\nabla^{2}+V(r,\theta)]\psi_{n}^{(m)}(\bm{r})=E_{n}\psi_{n}^{(m)}(\bm{r})
\end{equation}
where $\psi_{n}^{(m)}$ and $E_{n}^{(m)}$ are the eigenfunction and
eigenvalue, respectively.

Using single-center expansion for the electronic wavefunction
\begin{equation}
\psi_{n}^{(m)}(\bm{r})=\sum_{l=0}^{l_{max}}\frac{u_{nl}(r)}{r}Y_{lm}(\theta,\varphi)
\end{equation}
where $Y_{lm}(\theta,\varphi)$ are the spherical harmonics, the
radial wavefunction can be constructed with B-splines
\cite{Bachau2001}
\begin{equation}
u_{nl}(r)=\sum_{i=1}^{N_{l}}c_{il}^{n}B_{i}(r).
\end{equation}
Substituting Eqs.~(1), (14) and (15) into Eq.~(13)
and then projecting onto the $B_{i}Y_{lm}^{\ast}$
basis, we obtain the following matrix equation
\begin{equation}
HC=ESC
\end{equation}
where
\begin{eqnarray}
H_{il,i'l'}=&&\int_{0}^{r_{max}}\int_{0}^{\pi}\int_{0}^{2\pi}B_{i}(r)Y_{lm}^{\ast}(\theta,\varphi)
H_{el} \nonumber
\\ &&
B_{i'}Y_{l'm}(\theta,\varphi)dr \sin\theta d\theta d\varphi
\end{eqnarray}
\begin{equation}
S_{il,i'l'}=\delta_{ll'}\int_{0}^{r_{max}}B_{i}(r)B_{i'}(r)dr
\end{equation}
The eigenfunctions and eigenvalues  are obtained by diagonalizing Eq.~(16).

\subsection{Extracting asymptotic structure parameters}

In the asymptotic region, typically only a few terms in the
single-center expansion Eq.~(14) are important. Following Tong {\it
et al.} \cite{Tong2002}, we write the wavefunction of a linear
molecule as
\begin{equation}
\psi_{n}^{(m)}(\bm{r})=\sum_{l}F_{lm}(r)Y_{lm}(\theta,\varphi).
\end{equation}
In the MO-ADK theory \cite{Tong2002}, the radial functions in the asymptotic region
are fitted to the following form
\begin{equation}
F_{lm}(r)=C_{lm}r^{(Z_{c}/\kappa)-1}e^{-\kappa r}
\end{equation}
where $Z_{c}$  is the asymptotic charge, $\kappa=\sqrt{2I_{p}}$,
$I_{p}$ is the ionization energy.

\section{RESULTS AND DISCUSSION}
\subsection{On the quality of the model potential and the iteration procedure}
In this paper, the single-active-electron model potential [see
Sec.~II A] is created with the DFT, in which the exchange potential
is constructed with the exchange-only LDA potential and the LB model
potential (or LDA+LB). First we   check the quality of this model
potential if the molecular orbitals obtained from the standard
quantum chemistry package GAMESS \cite{Schmidt1993} are used as the
input.

\begin{figure}
\centering \mbox{{\myscaleboxa{
\includegraphics{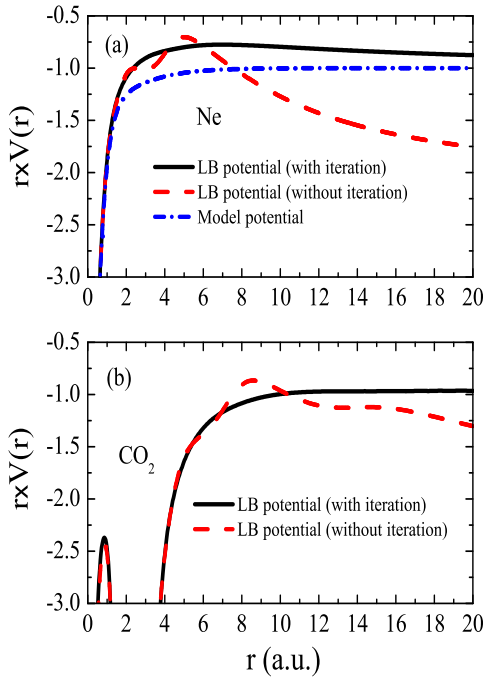}}}}
\caption{(Color online) (a) Effective charge of Ne with and without
the iteration (see text). Model potential is from \cite{Tong2005}.
(b) Effective charge of  CO$_2$ along the molecular axis. }
\end{figure}

In Fig.~1(a), we compare the present r-weighted LB  potential with
the empirical model potential of Tong {\it et al.} \cite{Tong2005}
for Ne. For clarity we plot the effective charge, defined as
$r\times V(r)$. The two potentials agree well in the small $r$
region. However, there are significant differences at large $r$. For
neutral atoms, the effective charge  should approach -1 at large
$r$. If the LB potential is calculated directly using the molecular
wavefunctions from GAMESS (dashed line) the effective charge
exhibits oscillations and then drops rapidly with $r$. This
undesirable behavior is due to the incorrect electron density, which
in turn is due to the limitation of the gaussian basis, calculated
from GAMESS in the large $r$ region. To correct this error, we
perform one more iteration on the potential: Firstly, an initial
model potential is generated using the Hartree-Fock (HF)
wavefunctions obtained from GAMESS. From this initial potential,
more accurate wavefunctions are obtained by solving Eq.~(13) with
B-spline basis. Then, a new model potential is constructed from
these new wavefunctions. From Fig.~1(a), we observe that the
effective charge obtained after one iteration (solid line) shows the
correct asymptotic behavior. The same procedure can be applied to
molecules. In Fig.~1(b), we show the model potential of CO$_{2}$
along the molecular axis, with and without  one iteration. It
confirms that the asymptotic behavior of the model potential is
correct after one iteration. We comment that in the case of CO$_2$
diffuse functions have been included in the basis sets. Clearly,
this alone is insufficient for obtaining accurate electron density
(or potential) at large $r$.

\subsection{Extracting molecular structure parameters for the MO-ADK theory}
Once the model potential is obtained, the eigenfunction and
eigenvalue can be calculated from solving Eq.~(13). In Table I,
  binding energies of rare gas atoms obtained using the present
method are compared to those from Ref.~\cite{Banerjee1999} and the
experimental values. Our method uses the same approximate exchange
potential as in Ref.~\cite{Banerjee1999}. The two calculations agree
in general, but small discrepancies do exist with experimental
values. These discrepancies can be reduced if correlation potential
is included in Eq.~(9). This fact has been well documented in
Ref.~\cite{Banerjee1999}.

\begin{table}
\caption{Comparison of calculated ionization energies of rare gas atoms in
the exchange-only LDA+LB model and experimental values.}
\begin{ruledtabular}
\begin{tabular}{c c c}
Atom & LDA+LB (a.u.)& I$_{p}$ (a.u.) \\
\hline
He& 0.786 &0.904$\footnote{Reference\cite{Emsley1998}}$ \\
  & 0.796$\footnote{Reference\cite{Banerjee1999}}$ & \\
Ne& 0.722 & 0.793$^{a}$ \\
  & 0.725$^{b}$ & \\
Ar& 0.524 & 0.579$^{a}$ \\
  & 0.528$^{b}$ & \\
Kr& 0.499 & 0.515$^{a}$ \\
Xe& 0.469 & 0.446$^{a}$ \\
\end{tabular}
\end{ruledtabular}
\end{table}

In Table II, we compare the ionization energies from the present
calculations with experimental vertical ionization energies for
several linear molecules. The equilibrium distances of these
molecules are also listed. The agreement between the calculated and
experimental values are good. Again we comment that the
exchange-only LDA+LB potential are used in our calculations. For
higher precision, correlation potential should be included
\cite{Schipper2000,Chu2005,Son2009,Telnov2009}.

\begin{table}
\caption{Equilibrium distances, ionization energies calculated  in the exchange-only LDA+LB model, and experimental vertical ionization potentials
for several linear molecules.}
\begin{ruledtabular}
\begin{tabular}{c c c c}
Molecule & R ({\AA}) & LDA+LB (eV) & I$_{p}$ (eV) \\
\hline
H$_{2}^{+}$& 1.058 & 29.99 & 29.99  \\
D$_{2}$& 0.742 & 13.65 & 15.47  \\
N$_{2}$& 1.098 & 14.99 & 15.58  \\
O$_{2}$& 1.208 & 10.62 & 12.03  \\
F$_{2}$& 1.412 & 16.03 & 15.70  \\
S$_{2}$& 1.889 & 10.36 & 9.36  \\
CO& 1.128 & 13.22 & 14.01  \\
NO& 1.151 & 9.14 & 9.26  \\
SO& 1.481 & 9.37 & 10.29  \\
CO$_{2}$& 1.163 & 14.63 & 13.78  \\
C$_{2}$H$_{2}$& 1.203 (R$_{CC}$) & 11.19 & 11.41  \\
&1.058 (R$_{CH}$)& &\\
HF& 0.917 & 15.03 & 15.77 \\
HCl& 1.275 & 11.41 & 12.75 \\
HCN& 1.067 (R$_{CH}$)& 13.46 & 13.80 \\
&1.159 (R$_{CN}$)& &\\
\end{tabular}
\end{ruledtabular}
\end{table}

\begin{table}
\caption{The newly fitted $C_{lm}$ coefficients vs values from earlier references, \cite{Tong2002,ATLe2007,Lin2006,Awasthi2008}.}
\begin{ruledtabular}
\begin{tabular}{c c c c c c c c c}
Molecule & $C_{0m}$ & $C_{1m}$ & $C_{2m}$&$C_{3m}$ & $C_{4m}$& $C_{5m}$ & $C_{6m}$&\\
\hline
H$_{2}^{+}$($\sigma_{g}$)& 4.52 & &0.62& &0.03&&&  \\
& 4.37 & &0.05& &0.00&&&\cite{Tong2002} \\
D$_{2}$($\sigma_{g}$)& 1.78& & 0.11 & &0.00&&&  \\
& 2.51 & &0.06& &0.00&& &\cite{Tong2002} \\
& 1.15 & &0.067& &0.001&& &\cite{Awasthi2008} \\
N$_{2}$($\sigma_{g}$)& 2.68& & 1.10 & &0.06&&&  \\
& 2.02 & &0.78& &0.04&& &\cite{Tong2002} \\
O$_{2}$($\pi_{g}$)& & & 0.52 & &0.03&&& \\
& & &0.62& &0.03&& &\cite{Tong2002} \\
F$_{2}$($\pi_{g}$)& & & 1.21 & &0.13&&&  \\
& & &1.17& &0.13&& &\cite{Tong2002} \\
S$_{2}$($\pi_{g}$)& & & 1.37 & &0.17&&&  \\
& & &0.81& &0.07&& &\cite{Tong2002} \\
CO($\sigma$)& 2.32&1.62 &0.82 & 0.17 &0.05&&&  \\
& 1.43&0.76 &0.28 & 0.02 &0.00&& &\cite{Tong2002} \\
NO($\pi$)& &0.21 &0.38 & 0.02 &0.02&&&  \\
& &0.22&0.41&0.01 & 0.00 && &\cite{Tong2002} \\
SO($\pi$)& &0.38 &0.71 & 0.05 &0.05&&&  \\
& &0.41&-0.31&0.01 & 0.00 && &\cite{Tong2002} \\
CO$_{2}$($\pi_{g}$)& & & 1.97 & &0.40&&0.04&  \\
& & &2.88& &1.71&&0.43 &\cite{ATLe2007} \\
C$_{2}$H$_{2}$($\pi_{u}$)& & 1.16 &  & 0.18&&0.02&& \\
& & 1.14 &  & 0.27&&0.04& &\cite{Lin2006}\\
HF($\pi$)& & 0.88 & 0.03 & 0.02& 0.01 &&& \\
HCl($\pi$)& & 1.23 & 0.01 & 0.05&0.01 &0.01&& \\
HCN($\pi$)& & 1.50 & 0.09 & 0.24&0.02 &0.02&& \\
\end{tabular}
\end{ruledtabular}
\end{table}

With the new wavefunctions, we  re-evaluate  the structure parameters for a number of linear molecules. Table III lists the newly fitted $C_{lm}$
coefficients with those listed in Tong {\it et al.} \cite{Tong2002} and in others, if available. These parameters will be used to obtain the
alignment-dependent tunneling ionization rates, following the MO-ADK theory \cite{Tong2002}.

\subsection{Comparison of alignment-dependent ionization probabilities between MO-ADK
 and other calculations}

Using the improved structure parameters tabulated in Table III, we
now use the analytical formula in Tong {\it et al.} \cite{Tong2002}
to obtain alignment-dependent tunneling ionization probabilities for
selected molecules that have also been carried out by other
 methods. The results are shown in Fig.~2. For simplicity, all the probabilities are normalized to
1.0 at the peak. First, we comment that for N$_{2}$, O$_{2}$,
F$_{2}$ the normalized probabilities obtained using the new
structure parameters do not show noticeable differences compared to
the probabilities calculated using old structure parameters. From
Table III, we note that the structure parameters for these three
molecules do not change much. We emphasize that in calculating
MO-ADK rates, one should always use the experimental vertical
ionization energy since the tunneling ionization rate depends
exponentially on the ionization potential. In Figs.~2(d) and 2(e),
we notice that, interestingly, the MO-ADK results using the new
$C_{lm}$ give stronger angular dependence than the old ones for both
H$_2^+$ and H$_2$. This is the result of the relatively larger
$C_{2m}$ as compared to $C_{0m}$ in the present calculations. For
H$_{2}^{+}$, the present result lies between the two calculations
from solving TDSE. For H$_{2}$, we compare the new results with
those from SFA, and the two agree quite well. For C$_{2}$H$_{2}$,
the new MO-ADK result agrees with the SFA, but differs from the
older MO-ADK \cite{Lin2006}. We comment that in the SFA calculation,
wavefunctions directly from the GAMESS code are used. In general,
SFA calculations yield incorrect total ionization rates.
Empirically, however, the normalized alignment dependence from the
SFA appears to be in agreement with the present MO-ADK. In
presenting the SFA results, we always use the renormalized ones. We
further comment that in SFA and other {\it ab initio} calculations,
ionization probability or rate for each alignment angle is
calculated independently. In the MO-ADK theory, the alignment
dependence is obtained analytically after the structure parameters
are obtained.

\begin{figure*}
\centering \mbox{{\myscaleboxb{
\includegraphics{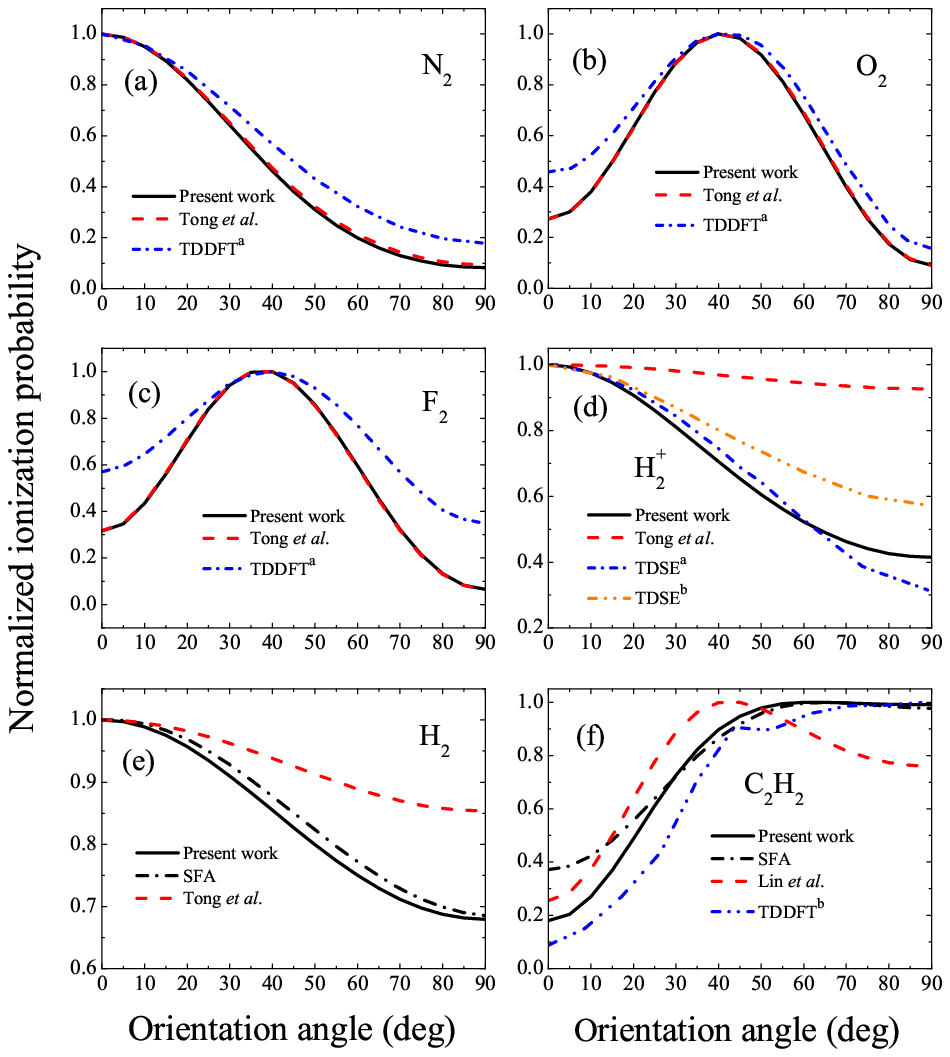}}}}
\caption{(Color online) Normalized alignment dependence of
ionization probability. (a) N$_{2}$ at laser intensity of $10^{14}$
W/cm$^{2}$; (b) O$_{2}$ at $10^{14}$ W/cm$^{2}$; (c) F$_{2}$ at
$2\times10^{14}$ W/cm$^{2}$; (d) H$_{2}^{+}$ at $5\times10^{14} $
W/cm$^{2}$; (e) H$_{2}$ at $2.3\times10^{14}$ W/cm$^{2}$; (f)
C$_{2}$H$_{2}$ at $5\times10^{13}$ W/cm$^{2}$. TDDFT$^{a}$ from
 Telnov {\it et al.} \cite{Telnov2009}, TDDFT$^{b}$ from  Otobe {\it et al.} \cite{Otobe2007}, TDSE$^{a}$ from Kamta {\it et al.} \cite{Kamta2006},
TDSE$^{b}$ from  Kjeldsen {\it et al.} \cite{Kjeldsen2007}, Tong
{\it et al.} from \cite{Tong2002} and Lin {\it et al.} from
\cite{Lin2006}.}
\end{figure*}

In recent years, {\it ab initio} calculations of molecular ionization by
intense lasers have been carried out by solving the TDDFT \cite{Telnov2009,Son2009,Otobe2007}. These
 calculations include all the electrons in the molecule.
Comparing to MO-ADK, in general, these calculations tend to give larger probabilities at angles where the ionization is small, see N$_{2}$ near
90$^{\circ}$ and O$_{2}$ and F$_{2}$ at angles near 0$^{\circ}$ and 90$^{\circ}$. For C$_{2}$H$_{2}$, on the other hand, the TDDFT result is
smaller at smaller angles than the present one. For this system, it was carried out by a different group \cite{Otobe2007}. Based on these results
we can say that the alignment dependence of the ionization probabilities obtained from MO-ADK and from TDDFT are in reasonable agreement. However,
we mention that  probabilities in Fig.~2 from MO-ADK include  ionization from the HOMO only,   while the many-electron TDDFT calculations show
significant contributions from the inner orbitals. More on the comparison between MO-ADK and TDDFT will be given later.

\subsection{Alignment dependence of ionization rates from HOMO, HOMO-1 and HOMO-2 orbitals}

Recently, strong field ionization phenomena involving inner orbitals
of molecules have been reported widely
\cite{McFarland2008,Smirnova2009,ATLe2009,Znakovskaya2009,Akagi2009}.
This is somewhat surprising since tunneling ionization rate
decreases very rapidly with the increase of ionization potential.
However, molecular tunneling ionization rates depend on the symmetry
of the orbital wavefunctions. For alignment angles where
$P(\theta)$ is near the minimum for the HOMO but where HOMO-1 is
near the maximum, there is a good possibility that ionization from
HOMO-1 can become comparable or higher than from HOMO. Indeed,
contribution from HOMO-1 to high-order harmonic generation (HHG)
from N$_{2}$ molecules has been reported by McFarland {\it et al.}
\cite{McFarland2008} when the molecules are aligned perpendicular to
the polarization of the probe laser. Le {\it et al.} \cite{ATLe2009}
have successfully reproduced the experimental results by including
HHG from HOMO and HOMO-1. Since tunneling ionization is the first
step for all rescattering processes
\cite{Toru08,ATLe09,Chen09,Micheau09}, including HHG \cite{ATLe09},
it is pertinent to investigate P($\theta$) from inner orbitals as
well.

\begin{table}
\caption{Comparison of calculated binding energies of HOMO, HOMO-1 and HOMO-2 of N$_{2}$, O$_{2}$ and CO$_{2}$ in the present exchange-only LDA+LB
model. Those from the LB$\alpha$ model and experimental vertical ionization potential are also given. Energies are in electron volts. For CO, HCl
and C$_{2}$H$_{2}$, only the energies of HOMO and HOMO-1 are considered.}
\begin{ruledtabular}
\begin{tabular}{c c c c c}
Molecule&Spin orbital&LDA+LB&LB$\alpha$&I$_{p}$ \\
\hline
N$_{2}$&3$\sigma_{g}$(HOMO)&15.0&15.5$\footnote{Reference \cite{Telnov2009}}$&15.6$\footnote{Reference \cite{Lofthus1977}}$\\
&1$\pi_{u}$(HOMO-1)&16.5&16.9$^{a}$&17.2$^{b}$\\
&2$\sigma_{u}$(HOMO-2)&17.8&18.5$^{a}$&18.7$^{b}$\\
O$_{2}$&1$\pi_{g}$(HOMO)&10.6&12.8$^{a}$&12.3$\footnote{Reference \cite{Baltzer1992}}$\\
&1$\pi_{u}$(HOMO-1)&17.3&17.4$^{a}$&16.7$^{c}$\\
&3$\sigma_{g}$(HOMO-2)&17.1&18.3$^{a}$&18.2$^{c}$\\
CO$_{2}$&1$\pi_{g}$(HOMO)&14.6&13.9$\footnote{Reference \cite{Son2009}}$&13.8$\footnote{Reference \cite{Turner1970}}$\\
&1$\pi_{u}$(HOMO-1)&18.3&17.5$^{d}$&17.6$^{e}$\\
&3$\sigma_{u}$(HOMO-2)&16.8&17.2$^{d}$&18.1$^{e}$\\
CO&5$\sigma$(HOMO)&13.2&&14.0$^{e}$\\
&1$\pi$(HOMO-1)&16.6&&16.9$^{e}$\\
HCl&2$\pi$(HOMO)&11.4&&12.8$\footnote{Reference \cite{Natalis1982}}$\\
&5$\sigma$(HOMO-1)&15.0&&16.3$^{f}$\\
C$_{2}$H$_{2}$&1$\pi_{u}$(HOMO)&11.2&&11.4$^{e}$\\
&3$\sigma_{g}$(HOMO-1)&15.7&&16.4$^{e}$\\
\end{tabular}
\end{ruledtabular}
\end{table}

In Table IV, the binding energies of HOMO, HOMO-1 and HOMO-2 for
several molecules are shown. These energies are compared to
calculations using the  LB$\alpha$ model and experimental values, to
check the relative accuracy of the model we have used. We emphasize
again that accurate experimental ionization energies, not the
theoretical values in the Table, are used in calculating the MO-ADK
rates. The extracted $C_{lm}$ parameters are given in Table V. Using
these parameters and experimental ionization energies, the alignment
dependence of ionization rates from different orbitals at a given
peak laser intensity can be readily calculated.

\begin{table}
\caption{The $C_{l}$ coefficients of HOMO, HOMO-1 and HOMO-2 for N$_{2}$, O$_{2}$ and CO$_{2}$ and of HOMO, HOMO-1 for CO, HCl and C$_{2}$H$_{2}$.
For $\sigma$ orbital, m=0 and $\pi$ orbital, m=1.}
\begin{ruledtabular}
\begin{tabular}{c c c c c c c}
Molecule&Spin orbital&&&$C_{l}$&& \\
\hline
N$_{2}$&&$C_{0m}$&$C_{2m}$&$C_{4m}$&&\\
&3$\sigma_{g}$(HOMO)&2.68&1.10&0.06&&\\
&&$C_{1m}$&$C_{3m}$&$C_{5m}$&&\\
&1$\pi_{u}$(HOMO-1)&1.89&0.22&0.01&&\\
&2$\sigma_{u}$(HOMO-2)&3.72&0.34&0.01&&\\
O$_{2}$&&&$C_{2m}$&$C_{4m}$&&\\
&1$\pi_{g}$(HOMO)&&0.52&0.03&&\\
&&$C_{1m}$&$C_{3m}$&$C_{5m}$&&\\
&1$\pi_{u}$(HOMO-1)&2.04&0.33&0.01&&\\
&&$C_{0m}$&$C_{2m}$&$C_{4m}$&&\\
&3$\sigma_{g}$(HOMO-2)&3.05&1.59&0.08&&\\
CO$_{2}$&&$C_{2m}$&$C_{4m}$&$C_{6m}$&&\\
&1$\pi_{g}$(HOMO)&1.97&0.40&0.04&&\\
&&$C_{1m}$&$C_{3m}$&$C_{5m}$&$C_{7m}$&\\
&1$\pi_{u}$(HOMO-1)&3.33&1.31&0.18&0.02&\\
&3$\sigma_{u}$(HOMO-2)&7.50&2.58&0.32&0.03&\\
CO&&$C_{0m}$&$C_{1m}$&$C_{2m}$&$C_{3m}$&$C_{4m}$\\
&5$\sigma$(HOMO)&2.32&1.62&0.82&0.17&0.05\\
&&$C_{1m}$&$C_{2m}$&$C_{3m}$&$C_{4m}$&$C_{5m}$\\
&1$\pi$(HOMO-1)&1.73&0.14&0.21&0.02&0.02\\
HCl&&$C_{1m}$&$C_{2m}$&$C_{3m}$&$C_{4m}$&$C_{5m}$\\
&2$\pi$(HOMO)&1.23&0.01&0.05&0.01&0.01\\
&&$C_{0m}$&$C_{1m}$&$C_{2m}$&$C_{3m}$&$C_{4m}$\\
&5$\sigma$(HOMO-1)&0.10&2.64&0.57&0.25&0.09\\
C$_{2}$H$_{2}$&&$C_{1m}$&$C_{3m}$&$C_{5m}$&&\\
&1$\pi_{u}$(HOMO)&1.16&0.18&0.02&&\\
&&$C_{0m}$&$C_{2m}$&$C_{4m}$&$C_{6m}$&\\
&3$\sigma_{g}$(HOMO-1)&4.40&3.85&0.72&0.09&\\
\end{tabular}
\end{ruledtabular}
\end{table}

\begin{figure*}
\centering \mbox{{\myscaleboxb{
\includegraphics{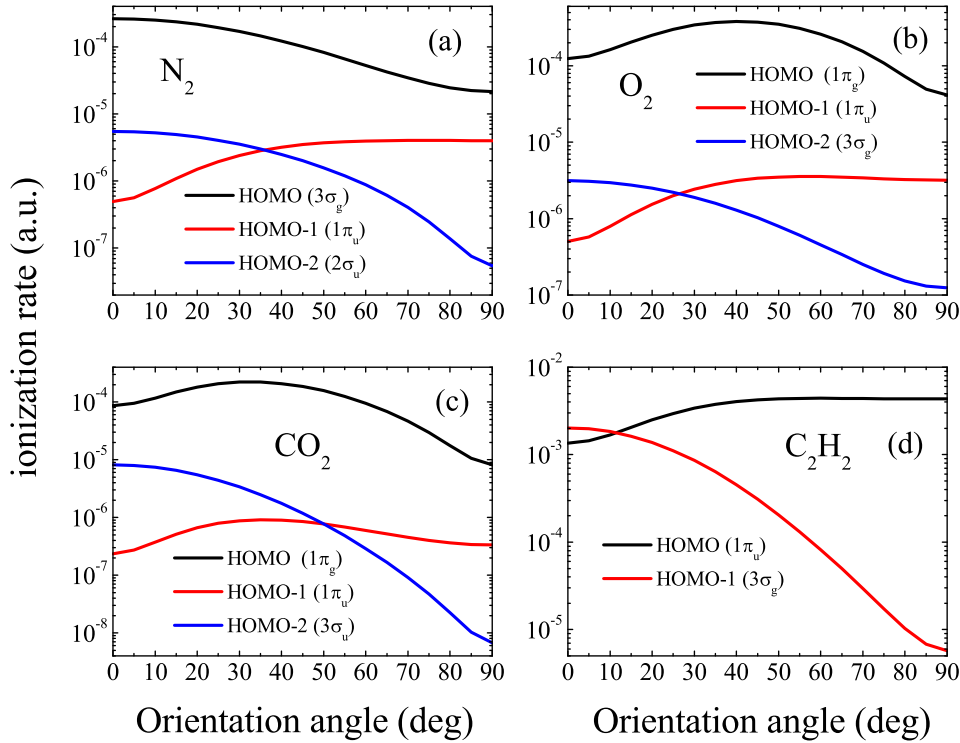}}}}
\caption{(Color online) Alignment dependence of ionization rates of
HOMO, HOMO-1 and HOMO-2 for N$_{2}$, O$_{2}$ and CO$_{2}$ and of
HOMO and HOMO-1 for C$_{2}$H$_{2}$. (a) N$_{2}$ at laser intensity
of $1.5\times10^{14}$ W/cm$^{2}$; (b) O$_{2}$ at $1.3\times10^{14}$
W/cm$^{2}$; (c) CO$_{2}$ at $1.1\times10^{14}$ W/cm$^{2}$; (d)
C$_{2}$H$_{2}$ at
  $1.5\times10^{14}$ W/cm$^{2}$.}
\end{figure*}

In Fig.~3, we compare the ionization rates from N$_{2}$, O$_{2}$ and
CO$_{2}$ molecules, for the HOMO, HOMO-1 and HOMO-2 orbitals, at
peak intensities indicated in the figure. Note that the angular
dependence, $P(\theta)$, reflects the symmetry of the molecular
orbital quite accurately. Thus a $\sigma$ orbital tends to have the
peak at 0$^{\circ}$ and a minimum at 90$^{\circ}$,  a $\pi_{g}$
orbital has the peak near 45$^{\circ}$ and minimum at 0$^{\circ}$
and 90$^{\circ}$, and a $\pi_{u}$ orbital has a peak near
90$^{\circ}$ and minimum near 0$^{\circ}$ (Deviations do occur, see
the HOMO-1 of CO$_2$ in Fig. 3(c)). These general behaviors of
ionization rates explain why HOMO-2 is bigger than HOMO-1 at small
angles for N$_{2}$, O$_{2}$, and CO$_{2}$, and why HOMO-1 is more
important than HOMO at small angles for C$_2$H$_2$. Note that the
relative ionization rates depend on laser intensities. The relative
ionization rates for inner orbitals increases faster with increasing
laser intensities. Using the parameters in Table V, their relative
rates can be easily calculated using the MO-ADK model. We have also
calculated the ionization rates using the molecular SFA. The
relative alignment dependence from SFA in general agrees with those
shown in Fig.~3. This is consistent with the findings in Le {\it et
al.} \cite{ATLe2009}.

\begin{figure*}
\centering \mbox{{\myscaleboxb{
\includegraphics{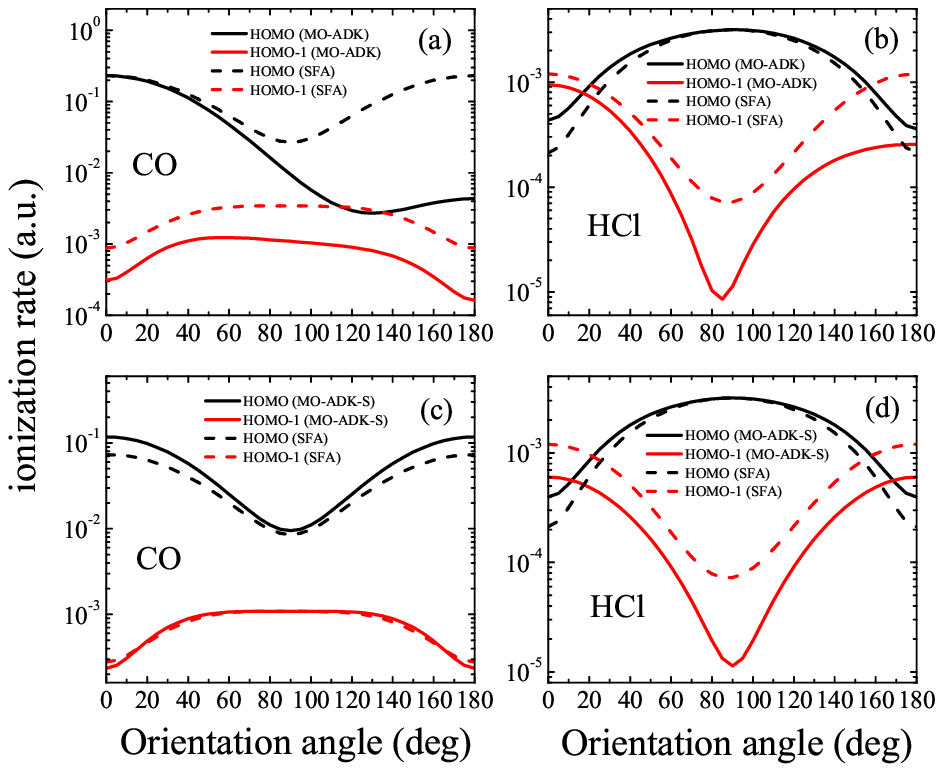}}}}
\caption{(Color online) Alignment dependence of ionization rates of
HOMO and HOMO-1. (a) CO at laser intensity of $4\times10^{14}$
W/cm$^{2}$; (b) HCl at $2\times10^{14}$ W/cm$^{2}$; (c) CO at
$4\times10^{14}$ W/cm$^{2}$; (d) HCl at $2\times10^{14}$ W/cm$^{2}$.
MO-ADK-S is the averaged MO-ADK rate for angles $\theta$ and
$\pi-\theta$.}
\end{figure*}

Fig.~4 shows the HOMO and HOMO-1 ionization rates for asymmetric diatomic molecules CO and HCl. There are recent experiments and other theoretical
calculations available for these two molecules \cite{Znakovskaya2009,Akagi2009}. For both systems, the predictions from MO-ADK are also compared
to results from SFA. Refer to Table IV, we note that the difference in binding energies between HOMO and HOMO-1 in CO is 2.9 eV, and 3.5 eV for
HCl. First we examine the $\theta$-dependence predicted by MO-ADK in Figs.~4(a) and 4(b). The HOMO of CO is a $\sigma$ orbital, its $P(\theta)$
drops rapidly from 0$^{\circ}$ to 90$^{\circ}$ and stays relative flat at larger angles. The HOMO-1 is a $\pi$ orbital and its $P(\theta)$ peaks
near 90$^{\circ}$. For HCl, the HOMO is a $\pi$ orbital and it peaks near 90$^{\circ}$. For the HOMO-1, it is a $\sigma$ orbital and its
P($\theta$) drops steadily till near 90$^{\circ}$. Interestingly, its $P(\theta)$ increases rapidly from 90$^{\circ}$ to 180$^{\circ}$, making it
almost like a symmetric molecule.

Why are the  $\sigma$ orbitals of the two molecules so different? It is due to the degree of asymmetry in the wavefunctions. Such asymmetry is
reflected in the $C_{l}$ coefficients in Table V. For CO (HCl), the first three coefficients are 2.32, 1.62, 0.82 (0.10, 2.64, 0.57) for $l$=0, 1
and 2, respectively. For HCl, there is one dominant $l$=1 component only, thus the ionization rate is nearly symmetric. For CO, the two
coefficients for $l$=0 and 1 are comparable, the wavefunction along the axis for $\theta=0$ and $\theta=\pi$ has the ratio
(2.32+1.62)/(2.32-1.62)=5.6. This gives a ionization rate ratio of 32, close to the value 50 read off from Fig. 4(a).

In Figs.~4(a) and 4(b), the $\theta$-dependence from SFA is
different from MO-ADK. Recall that in MO-ADK, static ionization rate
was calculated, thus a molecule is AB or BA with respect to the
fixed electric field will have different rates. For a linearly
polarized laser pulse, the direction of the electric field changes
after each half cycle, thus the cycle-averaged rates for AB and BA
are identical. To compare the SFA rate with the MO-ADK rate at an
angle $\theta$, we have to average the rates from the latter at
$\theta$ and $\pi$-$\theta$. These ``symmetrized" ionization rates
are denoted by MO-ADK-S in Fig.~4. By comparing the rates from SFA
and MO-ADK-S, we found in Fig.~4(c) that the two models agree well
for CO. For HCl, the relative rates for HOMO-1, normalized to HOMO,
are about a factor of two larger from SFA than from MO-ADK. We
comment that if ionization is measured using circularly polarized
light, the static MO-ADK rate can be compared directly with the rate
calculated using SFA.

The results of Figs.~3 and 4 show that at alignment angles where
tunneling ionization from the HOMO is large, contributions from
HOMO-1 or other inner orbitals are negligible. At alignment angles
where HOMO is near the  minimum, if the HOMO-1 (or even HOMO-2) is
near the maximum, then these inner orbitals may become important.
Since the relative tunneling ionization rates also depend on the
peak laser intensity, when multiple orbitals contribute to strong
field phenomena, the intensity dependence may become prominent.
Experimentally, such multiple orbital effects have been observed in
HHG from N$_2$ when molecules are aligned perpendicular to the
laser's polarization axis \cite{McFarland2008,ATLe2009}. The inner
orbitals have been shown to become important in HHG from CO$_2$ when
the molecules are aligned parallel to the laser's polarization axis
\cite{Smirnova2009}. Comparing to single photon ionization, strong
field ionization tends to be more selective by ionizing the HOMO.
For single photon ionization, cross sections for HOMO, HOMO-1 and
HOMO-2 in general have comparable values and often cross sections
from inner orbitals are higher, see e.g., \cite{Lucchese82} for
CO$_2$.

There are few theoretical alignment-dependent ionization rates from
inner orbitals available to compare with the predictions of the
MO-ADK theory presented here. There is an exception, however,
CO$_2$. In \cite{Spanner09} ionization rates from HOMO, HOMO-1 and
HOMO-2 have been calculated starting from the multielectron
perspective. In Fig.~5(a) we compare the rates from MO-ADK with
those from \cite{Spanner09} at the uncoupled channel approximation.
The two sets of calculations are normalized at the peak of the HOMO
curve. We note that the $\theta$-dependence agrees well for each
orbital. For the HOMO, the agreement is ``perfect". The rates for
the inner orbitals are larger from \cite{Spanner09} than from
MO-ADK. Part of the reason of the larger difference in the HOMO-1
rate could be due to the difference in the ionization energy used.
In \cite{Spanner09}, the energy difference between HOMO-1 and HOMO
was taken to be 3.53 eV, while in MO-ADK, the difference was taken
to be 3.80 eV from the experimental values in Table IV. For the
HOMO-2 the energy used is the same for the two calculations. The
alignment dependence of ionization rates for the three orbitals have
also been calculated in \cite{Smirnova2009} and the comparison with
the present MO-ADK is given in Fig.~5(b), again by normalizing at
the peak value of the HOMO.   In this case the differences are
larger. In \cite{Smirnova2009}, the ionization rates were calculated
using Coulomb corrected SFA plus sub-cycle dynamics. The TDDFT
method has also been used to obtain ionization probabilities from
different orbitals \cite{Son2009,Telnov2009}. For CO$_2$, the
predicted alignment dependence for the HOMO, HOMO-1 and HOMO-2 as
shown in Fig.~3 of \cite{Son2009} do agree with the present
Fig.~3(c), including that the peak for HOMO-1 is not at 90$^\circ$.
However, we should comment that in N$_2$ and O$_2$, the alignment
dependence using the same TDDFT method in \cite{Telnov2009} does not
agree with Figs.~3(a) and 3(b) shown for these two molecules.

\begin{figure*}
\centering \mbox{{\myscaleboxd{
\includegraphics{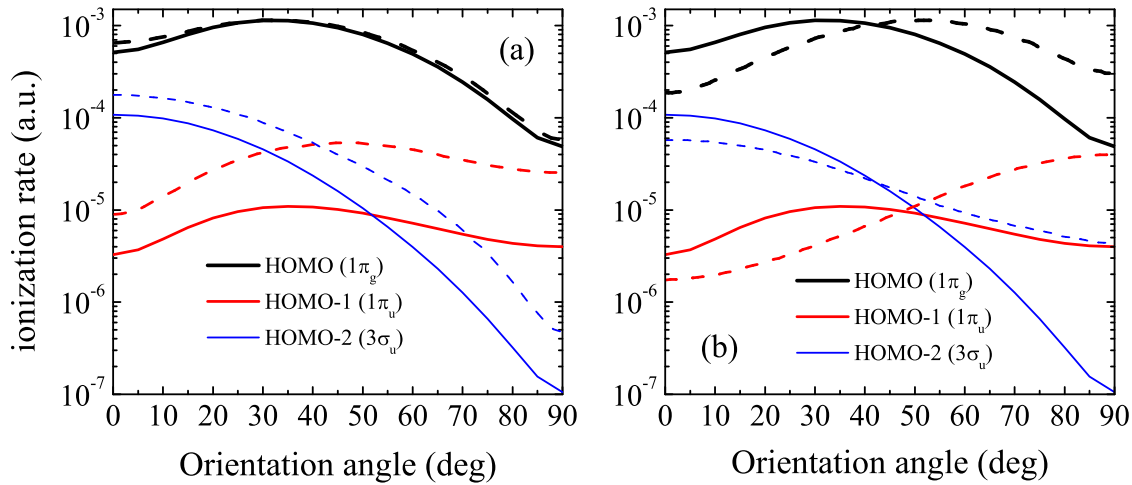}}}}
\caption{(Color online) Comparison of ionization rates of HOMO,
HOMO-1 and HOMO-2 of CO$_{2}$ at peak laser intensity of
$1.5\times10^{14}$ W/cm$^{2}$. The solid lines are from MO-ADK and
the dashed lines are from Spanner and Patchkovskii \cite{Spanner09}
(a) and from Smirnova {\it et al.} \cite{Smirnova2009} (b).}
\end{figure*}

\subsection{Comparisons with Experiments}

\begin{figure*}
\centering \mbox{{\myscaleboxb{
\includegraphics{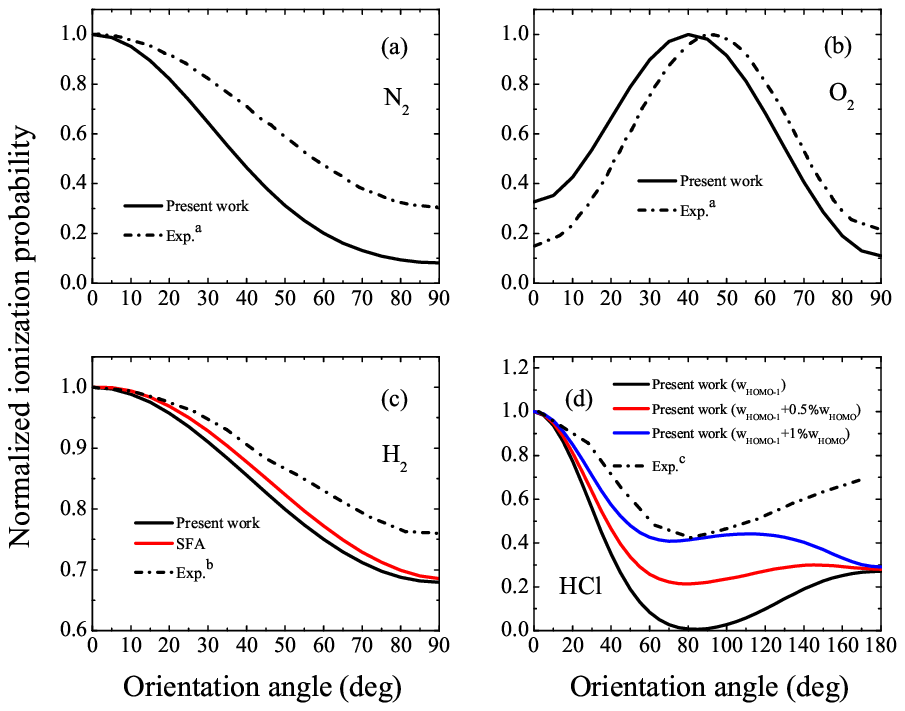}}}}
\caption{(Color online) Normalized alignment dependence of
ionization probability. (a) N$_{2}$ at laser intensity of
$1.5\times10^{14}$ W/cm$^{2}$; (b) O$_{2}$ at $1.3\times10^{14}$
W/cm$^{2}$; (c) H$_{2}$ at $2.3\times10^{14}$ W/cm$^{2}$; (d) HCl at
$1.4\times10^{14}$ W/cm$^{2}$. Linearly polarized lights for (a) and
(b); Circularly polarized lights for (c) and (d). Exp.$^{a}$ from
Pavi\v{c}i\'{c} {\it et al.} \cite{Pavicic2007}; Exp.$^{b}$ from
Staudte {\it et al.} \cite{Staudte2009} and Exp.$^{c}$ from  Akagi
{\it et al.} \cite{Akagi2009}. Additional symbols for (d), see
text.}
\end{figure*}

Fig.~6 shows the normalized alignment dependence of ionization
probability of N$_{2}$, O$_{2}$, H$_{2}$ and HCl. From Figs.~6(a)
and 6(b), we comment that the normalized ionization probability of
N$_{2}$ calculated from MO-ADK theory using the old and the newly
fitted coefficients agree quite well (no visible difference in the
plot). Compared to the experiment of Pavi\v{c}i\'{c} {\it et al.}
\cite{Pavicic2007}, the MO-ADK theory shows some differences. But
the difference is considered acceptable. Note that the determination
of alignment dependence from the experiment has angular average
which was not included in the theory curve. Take the experimental
result as reference, the TDDFT result (see Fig.~2) is better than
the MO-ADK for N$_{2}$. For O$_{2}$, it is the other way around. The
same comparison for CO$_{2}$ has been addressed in an earlier paper
\cite{Zhao2009}. In that case, the old MO-ADK results were found to
be inaccurate due to the inaccuracy of the old $C_{l}$ parameters.
In \cite{Zhao2009} it was further concluded that the experimental
P($\theta$) from \cite{Pavicic2007} appears to be too narrowly
peaked. We note that the new result from \cite{Spanner09} also does
not agree with the experiment. However the authors suspect that the
discrepancy is due to intermediate excitation channels were not
included in their calculation. We tend to think that additional
experiments are needed to help resolving this discrepancy.

In Fig.~2(e) we show that the MO-ADK probabilities for H$_{2}$ using
the new structure parameters are different from using the earlier
ones \cite{Tong2002}. The new MO-ADK probabilities and molecular SFA
agree well, see Fig.~6(c). Comparing to experimental data of Staudte
{\it et al.}~\cite{Staudte2009}, the agreement is good in view that
the theory curve has not included average over angular resolution.
In \cite{Staudte2009}, the ratio of ionization rate for molecules
aligned parallel vs perpendicular, with respect to the polarization
axis, were also determined at four intensities from 2 to
4.5$\times$10$^{14}$ W/cm$^2$ (for circularly polarized laser). The
ratio from the present SFA (not shown) agree with the SFA model in
that paper, and with the new MO-ADK ratio of 1.45 (the old MO-ADK
gives 1.15). We expect the theoretical ratio be reduced somewhat if
angular average is incorporated. We mention that a similar
measurement at one intensity for laser wavelength of 1850 nm was
reported in \cite{Magrakvelidze2009}, which gives a ratio of 1.15.
Interestingly, this ratio was reported to be 3.0 \cite{Hoff2009} in
another recent experiment, while the theory presented in the same
paper gives a ratio of 2.1. We comment that the ratio is taken at
the maximum with respect to the minimum and thus sensitive to the
angular average. Comparison of the rates over the whole angular
range would be preferable.

In Fig.~6(d), the $P(\theta)$ of the HOMO-1 orbital in HCl reported
in Ref.~\cite{Akagi2009} using circularly polarized light at the
intensity of 1.4$\times$10$^{14}$ W/cm$^2$ is shown. We compare the
HOMO-1 result from the MO-ADK theory using the laser parameters
given in the experiment, and by normalizing the data at
$\theta$=0$^{\circ}$. In Ref.~\cite{Akagi2009}, the alignment
dependence for HOMO and HOMO-1 has also been reported using the
TDDFT. The alignment dependence between MO-ADK and TDDFT
calculations are quite similar, but our relative HOMO-1 probability
is about a factor of three higher at the same laser intensity. The
ionization probability from both calculations drop much faster from
0$^{\circ}$ to 90$^{\circ}$ when compared to the experiment. By
introducing a small fraction of the contribution from the HOMO in
the manner suggested in \cite{Akagi2009}, the MO-ADK theory can
achieve a reasonable agreement with the experimental data from
0$^{\circ}$ to 90$^{\circ}$, see Fig. 6(d). On the other hand, the
agreement at angles larger than 90$^{\circ}$ is still not as good.

\subsection{Ionization probability of H$_{2}^{+}$ }

\begin{figure*}
\centering \mbox{{\myscaleboxb{
\includegraphics{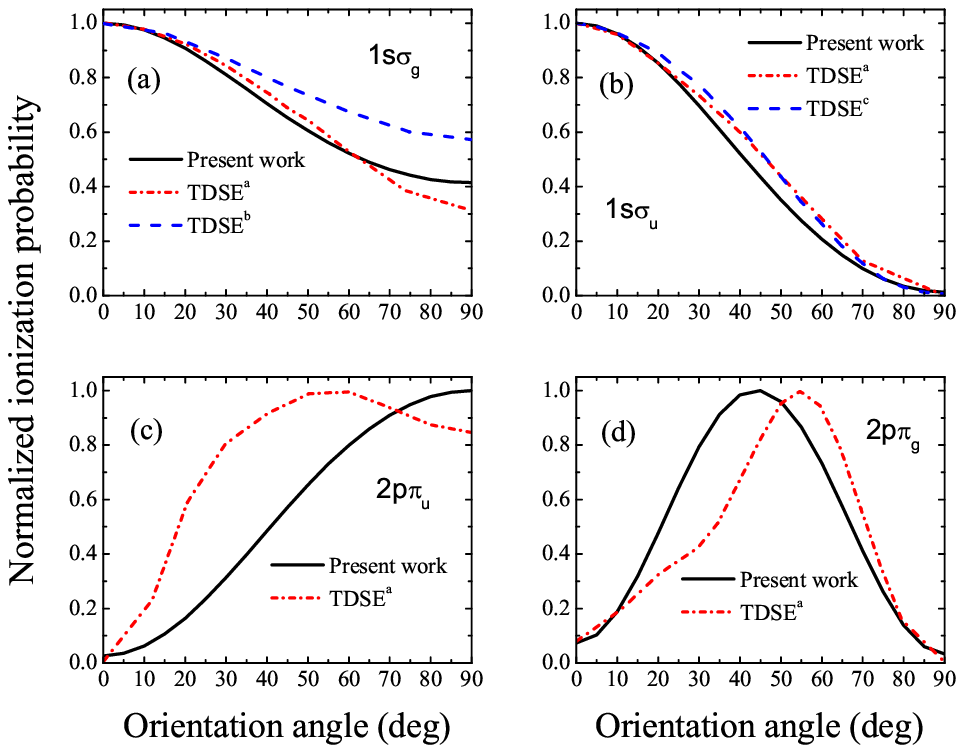}}}}
\caption{(Color online) Normalized alignment dependence of
ionization probability of H$_{2}^{+}$. (a) 1s$\sigma_{g}$ at laser
intensity of $5\times10^{14}$ W/cm$^{2}$; (b) 1s$\sigma_{u}$ at
  $10^{14}$ W/cm$^{2}$; (c) 2p$\pi_{u}$ at
  $10^{13}$ W/cm$^{2}$; (d) 2p$\pi_{g}$ at
$10^{12}$ W/cm$^{2}$. TDSE$^{a}$ from  Kamta {\it et al.}
\cite{Kamta2006}, TDSE$^{b}$ from  Kjeldsen {\it et al.}
\cite{Kjeldsen2007} and TDSE$^{c}$ from Telnov {\it et al.}
\cite{Telnov2007}.}
\end{figure*}

\begin{figure}
\centering \mbox{{\myscaleboxc{
\includegraphics{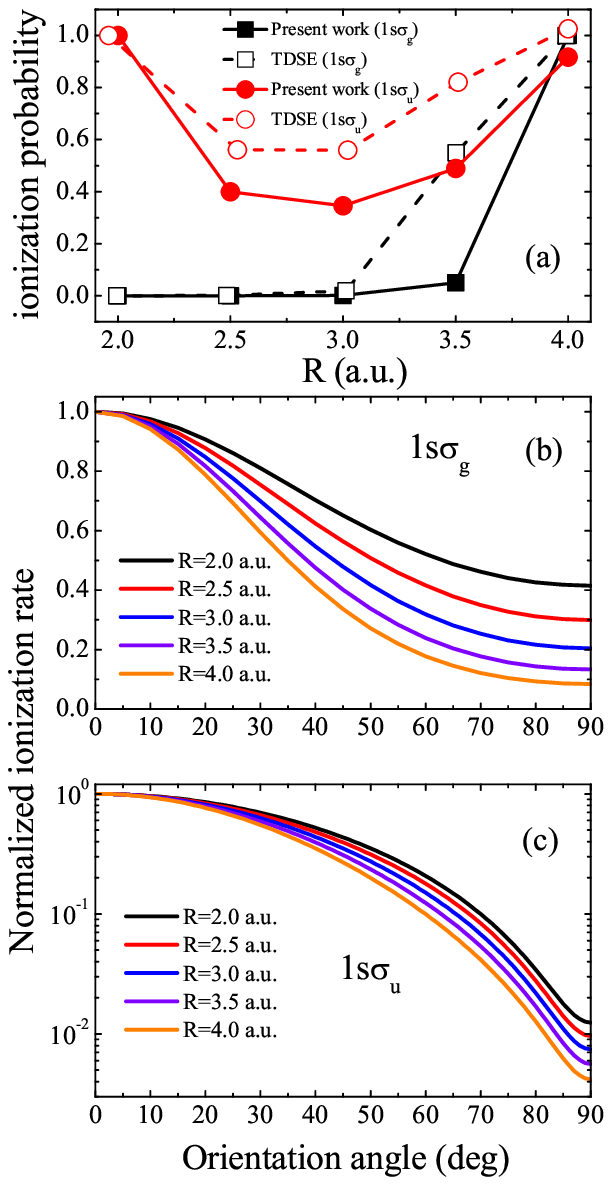}}}}
\caption{(Color online) (a) R-dependence of the normalized
ionization probability of H$_{2}^{+}$ at laser intensity of
$10^{14}$ W/cm$^{2}$; see text. (b) The normalized alignment
dependence of ionization rate at $5\times10^{14}$ W/cm$^{2}$ for
1s$\sigma_{g}$; (c) The normalized alignment dependence of
ionization rate at $10^{14}$ W/cm$^{2}$ for 1s$\sigma_{u}$. TDSE
from \cite {Kamta2007}.}
\end{figure}

The ionization probability of H$_{2}^{+}$ has been calculated from
solving the TDSE by different groups
\cite{Kjeldsen2007,Kamta2006,Telnov2007}. It is of interest to
compare the predictions based on MO-ADK with those from solving the
TDSE. In Fig.~7, the normalized alignment-dependent ionization
probability from the first four molecular orbitals of H$_{2}^{+}$ at
the equilibrium distance are shown. The data for 1s$\sigma_{g}$ have
been discussed earlier \cite{Zhao2009}. For ionization from
1s$\sigma_{u}$, the two TDSE calculations and the MO-ADK agree quite
well. For 2p$\pi_{u}$, the MO-ADK theory tends to peak at
90$^{\circ}$ while the TDSE result gives a peak closer to about
60$^{\circ}$. For 2p$\pi_{g}$ state, the MO-ADK predicts a peak near
45$^{\circ}$ while TDSE calculation gives a peak at about
55$^{\circ}$. Note different peak laser intensities are used for the
ionization from each orbital.

In Fig.~8(a) we show the dependence of normalized ionization
probabilities vs the internuclear separation  for the
1s$\sigma_{g,u}$ states of H$_{2}^{+}$ with the molecular axis
parallel to the polarization axis. The results are compared to the
TDSE calculations of \cite{Kamta2007}. By normalizing the
probability at $R=2~a.u.$  for the 1s$\sigma_{u}$, we find that
there is a general good agreement between the TDSE result and from
the MO-ADK. For the 1s$\sigma_g$, the two calculations are
normalized at $R=4.0~a.u.$. For both calculations, the probabilities
at $R$ less than $3.5~a.u.$ are significantly smaller than at
$R=4.0~a.u.$. In Figs.~8(b) and 8(c), the normalized
alignment-dependent ionization rates are shown for different $R$.
Clearly as $R$ increases, the angular dependence becomes sharper.
This is easily understood for $\sigma$ orbitals since the molecular
orbital becomes more elongated along the molecular axis as $R$
increases. The $C_{l}$ coefficients are tabulated in Table VI to
reflect how these parameters vary as $R$ increases.

\begin{table}
\caption{The $C_{l}$ coefficients of 1s$\sigma_{g}$, 1s$\sigma_{u}$
, 2p$\pi_{g}$ and 2p$\pi_{u}$ for H$_{2}^{+}$ at different
internuclear distances. For $\sigma$ orbital, m=0 and $\pi$ orbital,
m=1. The calculated binding energies and exact ones (in atomic
units) are also listed.}
\begin{ruledtabular}
\begin{tabular}{c c c c c c c}
symmetry&$R~(a.u.)$&&binding energy&&$C_{l}$& \\
\hline
1s$\sigma_{g}$&&Present&Exact&$C_{0m}$&$C_{2m}$&$C_{4m}$\\
&2.0&1.1025&1.1026&4.52&0.62&0.03 \\
&2.5&0.9937&0.9938&4.25&0.81&0.05 \\
&3.0&0.9107&0.9109&4.10&1.04&0.08 \\
&3.5&0.8463&0.8466&4.00&1.31&0.12 \\
&4.0&0.7958&0.7961&3.96&1.60&0.19 \\
1s$\sigma_{u}$&&&&$C_{1m}$&$C_{3m}$&$C_{5m}$\\
&2.0&0.6674&0.6675&1.89&0.08&0.00 \\
&2.5&0.6920&0.6921&2.18&0.15&0.00 \\
&3.0&0.7012&0.7014&2.48&0.25&0.01 \\
&3.5&0.7009&0.7012&2.79&0.37&0.02 \\
&4.0&0.6952&0.6956&3.13&0.53&0.04 \\
2p$\pi_{g}$&&&&&$C_{2m}$&$C_{4m}$\\
&2.0&0.4288&0.4288&&0.11&0.002 \\
2p$\pi_{u}$&&&&$C_{1m}$&$C_{3m}$&\\
&2.0&0.2267&0.2267&0.90&0.02& \\
\end{tabular}
\end{ruledtabular}
\end{table}

\section{CONCLUSIONS}
In this paper we proposed a new method to obtain accurate molecular wavefunctions in the asymptotic region starting with molecular orbitals
obtained from the widely used quantum chemistry packages such as GAMESS and GAUSSIAN. From these wavefunctions, the structure parameters in the
molecular tunneling ionization theory (MO-ADK) of Tong {\it et al.} \cite{Tong2002} can be accurately determined. Using these structure
parameters, we re-examined the alignment-dependent tunneling ionization probabilities for a number of molecules, including ionization from HOMO-1
and HOMO-2 orbitals. The calculated tunneling ionization probabilities are compared to probabilities determined from experiments, and to several
other more elaborated calculations. Since tunneling ionization is the first step for strong field phenomena involving molecular targets, these
structure parameters are useful and thus are tabulated. The procedure for obtaining the structure parameters discussed in this paper is generally
applicable to any linear molecules. Despite of its fundamental importance, accurate strong field alignment-dependent ionization probabilities are
still not widely available. Experimental measurements as well as more advanced calculations tend to deal with different molecules and under
different conditions, thus it is difficult to benchmark the accuracy of the theoretical models. While MO-ADK model is the simplest model for
obtaining tunneling ionization rates, it appears that its predictions so far are in good agreement with most of the experimental data and with
most the elaborate theoretical calculations.\\

\begin{acknowledgements}
This work was supported in part by Chemical Sciences, Geosciences
and Biosciences Division, Office of Basic Energy Sciences, Office of
Science, U.S. Department of Energy. S.-F.Z was also supported by the
National Natural Science Foundation of China under Grant
No.10674112.
\end{acknowledgements}

\end{document}